\documentclass[conference]{IEEEtran}
\IEEEoverridecommandlockouts
\usepackage{cite}
\usepackage{amsmath,amssymb,amsfonts}
\usepackage{algorithmic}
\usepackage{graphicx}
\usepackage{textcomp}
\usepackage{url}
\usepackage{multirow}
\usepackage{xcolor}
\usepackage{booktabs}
\usepackage{array}
\usepackage{enumitem}
\usepackage{subfigure}
\def\BibTeX{{\rm B\kern-.05em{\sc i\kern-.025em b}\kern-.08em
    T\kern-.1667em\lower.7ex\hbox{E}\kern-.125emX}}
\begin{document}

\title{When Bugs Linger: A Study of Anomalous Resolution Time Outliers and Their Themes}

\author{

\IEEEauthorblockN{Avinash Patil}
\IEEEauthorblockA{
\textit{Juniper Networks Inc.}\\
Sunnyvale, USA \\
ORCID: 0009-0002-6004-370X}

}

\maketitle

\begin{abstract}
Efficient bug resolution is critical for maintaining software quality and user satisfaction. However, specific bug reports experience unusually long resolution times, which may indicate underlying process inefficiencies or complex issues. This study presents a comprehensive analysis of bug resolution anomalies across seven prominent open-source repositories: \textit{Cassandra}, \textit{Firefox}, \textit{Hadoop}, \textit{HBase}, \textit{SeaMonkey}, \textit{Spark}, and \textit{Thunderbird}. Utilizing statistical methods such as Z-score and Interquartile Range (IQR), we identify anomalies in bug resolution durations. To understand the thematic nature of these anomalies, we apply Term Frequency-Inverse Document Frequency (TF-IDF) for textual feature extraction and KMeans clustering to group similar bug summaries. Our findings reveal consistent patterns across projects, with anomalies often clustering around test failures, enhancement requests, and user interface issues. This approach provides actionable insights for project maintainers to prioritize and effectively address long-standing bugs.
\end{abstract}


\section{Introduction}

Software development projects often rely on issue-tracking systems to manage bug reports, feature requests, and other tasks. Efficiently resolving these issues is critical to maintaining software quality and user satisfaction. However, some bug reports experience unusually long resolution times, which can indicate deeper problems such as test instability, unclear ownership, or architectural complexity.

Detecting such anomalies in bug resolution times can help project managers identify bottlenecks and allocate resources more effectively. Traditional approaches to anomaly detection in software engineering have included statistical methods and rule-based systems. For instance, Hangal and Lam introduced DIDUCE, a dynamic invariant detection tool that aids in identifying software errors by monitoring program executions \cite{hangal2002tracking}.

In recent years, machine learning and natural language processing (NLP) techniques have been applied to analyze bug reports. These methods can automatically categorize and cluster bug reports based on textual similarities, aiding in triage and prioritization. For example, Nguyen et al. proposed an unsupervised approach to categorize bug reports using clustering and labeling algorithms, enhancing the understanding of bug report topics \cite{limsettho2016unsupervised}.

Furthermore, studies have explored the use of NLP for predicting the fault category of software bugs. Lamkanfi et al. demonstrated that textual descriptions in bug reports could be leveraged to predict fault categories, improving the efficiency of bug triaging \cite{lamkanfi2010predicting}.

Despite these advancements, a need remains for comprehensive approaches that combine anomaly detection with resolution times and NLP-based clustering of bug report summaries. Such integrated methods can provide deeper insights into the nature of anomalies and support more informed decision-making in software maintenance.

In this study, we analyze bug report datasets from multiple open-source repositories to identify anomalies in bug resolution times and apply natural language processing (NLP) techniques to cluster anomalous bug summaries. Our goal is to identify patterns and themes associated with prolonged bug resolutions, thereby contributing to a deeper understanding and improvement of software maintenance processes.

\section{Related Work}

Anomaly detection and bug report analysis are critical areas in software engineering, with numerous studies exploring various methodologies to enhance software reliability and maintenance.

\subsection{Anomaly Detection in Software Systems}

Hangal and Lam introduced DIDUCE, a dynamic invariant detection tool that aids in identifying software bugs by monitoring program executions for anomalies~\cite{hangal2002tracking}. Their approach emphasizes the importance of runtime analysis in detecting deviations from expected behavior.

In the realm of time series data, statistical methods have been employed for anomaly detection. For instance, Booking.com detailed its approach to identifying anomalies in time series using statistical analysis, highlighting the effectiveness of such methods in real-world applications~\cite{booking2023anomaly}.

Recent advancements have integrated machine learning with anomaly detection frameworks. Moharam et al. presented an anomaly detection framework using Digital Twin (DT) technology to simulate and monitor dynamic radio environments. By modeling network conditions and attack scenarios, the DT enables accurate identification of anomalies, particularly security threats~\cite{moharam2025anomaly}.

\subsection{Clustering and Classification of Bug Reports}

Clustering techniques have been widely used to categorize and prioritize bug reports. Chen et al. proposed a graph partitioning algorithm to cluster defect reports, facilitating better understanding and management of software defects~\cite{chen2009clustering}. Similarly, Meng et al. developed an automatic classification method for bug reports that considers both textual information and the intent behind reports, improving classification performance~\cite{meng2022automatic}.

The application of TF-IDF and KMeans clustering has been explored in various contexts. For example, Becker demonstrated the use of these techniques in clustering U.S. laws, showcasing their versatility in text analysis tasks~\cite{becker2015clustering}.

A recent study by Patil et al. proposed a novel approach based on machine learning methods to automatically detect duplicate bug reports in large open bug repositories. Their approach demonstrates promising results across various methods ~\cite{patil2023auto}.

\subsection{Semantic Similarity and Embedding Models}

Understanding semantic similarity between bug reports is essential for effective deduplication and triaging. Patil et al. conducted a comparative study of text embedding models, including TF-IDF, FastText, and BERT, to assess their performance in capturing semantic similarities in bug reports~\cite{patil2024comparative, patil2025next}. Their findings suggest that advanced embedding models can significantly enhance the accuracy of bug report analysis.

\subsection{Bug Report Classification Techniques}

Terdchanakul et al. introduced a classification model utilizing N-gram IDF to distinguish between bug and non-bug reports, addressing the challenge of misclassification in bug tracking systems~\cite{terdchanakul2017bug}. Their approach highlights the significance of feature selection in enhancing classification accuracy.

A comprehensive analysis by Alshamrani et al. evaluated leading machine learning techniques for bug report classification, including classical ML, BERT-based techniques, and AutoML. Their study provides insights into the effectiveness of various approaches in accurately classifying bug reports~\cite{alshamrani2025ml}.

\subsection{Dependency-Aware Bug Triaging}

Efficient bug triaging is crucial for resolving bugs in a timely manner. Jahanshahi et al. proposed DABT, a dependency-aware bug triaging method that considers textual information, bug dependencies, and associated costs to assign bugs to appropriate developers, thereby reducing overdue bugs and average fixing time~\cite{jahanshahi2021dabt}.

Building upon this, they introduced S-DABT, a Schedule and Dependency-aware Bug Triage method that utilizes integer programming and machine learning techniques to assign bugs to suitable developers. This approach considers textual data, bug-fixing costs, bug dependencies, and developers' schedules, resulting in improved bug-fixing times and more efficient task distribution~\cite{jahanshahi2022s}.

These studies collectively inform our methodology, which integrates statistical anomaly detection with TF-IDF-based clustering to analyze bug resolution behaviors across multiple software repositories.

\section{Methodology}

This section outlines the approach used for detecting resolution time anomalies and analyzing their thematic content across seven open-source software repositories. The methodology includes four key components: data preprocessing, anomaly detection, textual feature extraction, and clustering.

\subsection{Dataset}

This study utilizes the \textit{GitBugs} dataset~\cite{patil2025gitbugs}, a curated collection of historical bug reports from multiple large-scale open-source projects. GitBugs standardizes metadata from platforms such as Apache JIRA and Bugzilla, providing a unified schema with fields including issue ID, creation and resolution timestamps, summary text, priority, and status.

We selected seven well-known repositories from the dataset—\textit{Cassandra}, \textit{Firefox}, \textit{Hadoop}, \textit{HBase}, \textit{SeaMonkey}, \textit{Spark}, and \textit{Thunderbird}—to ensure diversity in domain, scale, and bug tracking systems. These repositories range from infrastructure-level projects (e.g., Hadoop, Spark) to end-user applications (e.g., Firefox, Thunderbird), allowing for broad generalization of our findings.

Table~\ref{tab:gitbugs_summary} presents key statistics across these projects, including the total number of bug reports and the prevalence of duplicate issues, a common characteristic of large-scale bug tracking data.

\begin{table}[h]
\centering
\caption{Summary Statistics Across Projects in GitBugs}
\begin{tabular}{lrrr}
\hline
\textbf{Project} & \textbf{Total Reports} & \textbf{Duplicates} & \textbf{Duplicate Rate (\%)} \\
\hline
Cassandra & 4,612 & 300 & 6.5 \\
Firefox & 28,824 & 6,255 & 21.7 \\
Hadoop & 2,503 & 128 & 5.1 \\
HBase & 5,403 & 108 & 2.0 \\
SeaMonkey & 1,076 & 120 & 11.2 \\
Spark & 20,275 & 497 & 2.5 \\
Thunderbird & 15,192 & 4,200 & 27.6 \\
\hline
\end{tabular}
\label{tab:gitbugs_summary}
\end{table}

The dataset is publicly available at \url{https://github.com/av9ash/gitbugs}, providing a reproducible resource for empirical software engineering research.

\subsection{Data Preprocessing}

Each dataset consisted of bug reports with fields including issue ID, creation and resolution timestamps, priority, status, and textual summaries. We first converted all timestamp fields to DateTime format and calculated resolution duration as the number of days between the `Created` and `Resolved` fields. We excluded bugs without resolution dates from anomaly analysis.

\subsection{Anomaly Detection}

We applied two complementary statistical techniques to identify anomalously long resolution times:

\begin{itemize}
    \item \textbf{Z-Score Method}: For each bug, we calculated the z-score based on the distribution of resolution times. We flagged bugs with absolute z-scores greater than three as anomalies~\cite{zscore1998, openstax2022zscore}.
    \item \textbf{Interquartile Range (IQR) Method}: We considered bugs with resolution times below $Q_1 - 1.5 \times IQR$ or above $Q_3 + 1.5 \times IQR$ also as anomalous~\cite{tukey1977exploratory}.
\end{itemize}

We included bugs identified by either method in the anomaly set.

\subsection{Text Vectorization}

To analyze the textual content of anomalous bug reports, we applied TF-IDF (Term Frequency-Inverse Document Frequency)~\cite{salton1988term} to the bug summary field. TF-IDF transforms a corpus into a numerical matrix that reflects term relevance across the document set while down weighting common words.

\subsection{Dimensionality Reduction and Clustering}

To visualize and group similar bug summaries, we used Principal Component Analysis (PCA)~\cite{wold1987principal} to reduce the TF-IDF matrix to two dimensions. We then applied KMeans clustering~\cite{lloyd1982least} to discover thematic groupings within the anomalous summaries.

We empirically selected $k=3$ clusters based on prior literature suggesting this number balances granularity and interpretability in software defect categorization~\cite{herzig2013}. Clusters were interpreted based on their top TF-IDF-weighted terms.

We performed all analysis using Python and the \texttt{scikit-learn}~\cite{pedregosa2011scikit} and \texttt{matplotlib}~\cite{hunter2007matplotlib} libraries.

\section{Results}

To assess anomaly detection across diverse software ecosystems, we analyzed bug report datasets from seven major open-source repositories: \textit{Cassandra}, \textit{Firefox}, \textit{Hadoop}, \textit{HBase}, \textit{SeaMonkey}, \textit{Spark}, and \textit{Thunderbird}. We processed each dataset to extract resolution times and flag anomalous delays using Z-score and IQR-based methods. We further applied text clustering on anomalous bug summaries using TF-IDF and KMeans, revealing thematic groupings.

Due to space constraints, only the visualizations for \textit{Cassandra}, \textit{Firefox}, \textit{Hadoop}, and \textit{HBase} are shown below. Figures for \textit{SeaMonkey}, \textit{Spark}, and \textit{Thunderbird} have been moved to the Appendix.

\subsection{Cassandra}
The \textit{Cassandra} project showed a wide dispersion in resolution times, with anomalies extending well beyond 1500 days. Anomaly density was particularly high in the 2020–2022 range. Clustering of anomalous summaries revealed three distinct themes:
\begin{itemize}
    \item \textbf{Cluster 0:} Test instability — keywords included \textit{test}, \textit{failure}, \textit{flaky}.
    \item \textbf{Cluster 1:} Enhancement proposals — dominated by \textit{CEP}, \textit{add}, \textit{update}.
    \item \textbf{Cluster 2:} General issues referencing core packages like \textit{apache}, \textit{org}.
\end{itemize}

\begin{figure*}[ht]
    \centering
    \subfigure[Bug resolution times in Cassandra (anomalies in red).]{
        \includegraphics[width=0.3\textwidth]{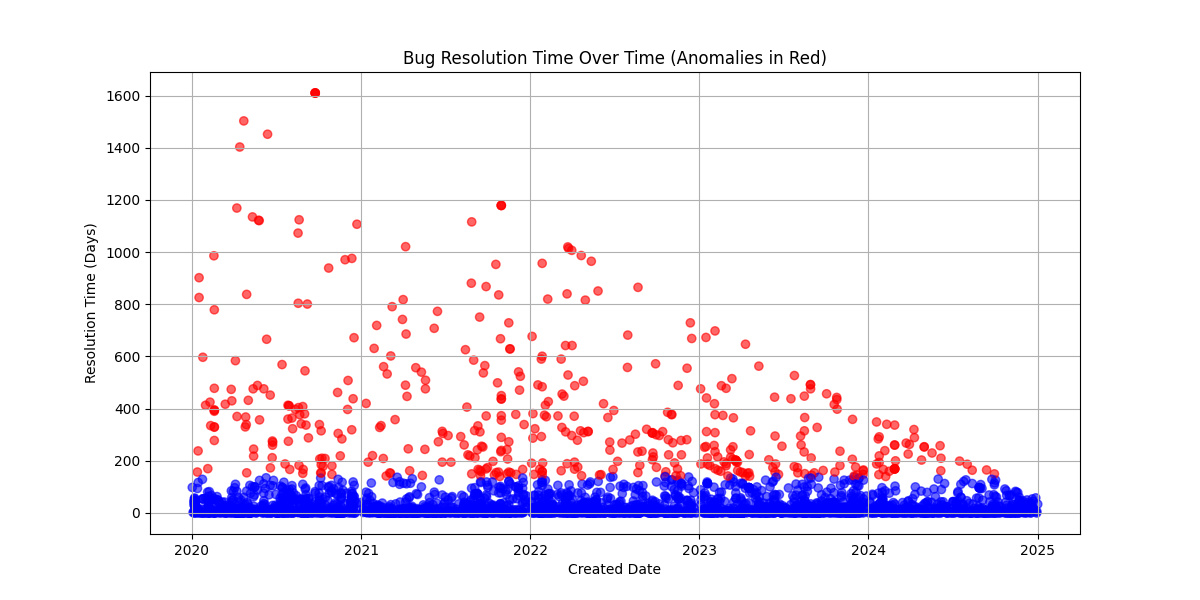}
        \label{fig:cassandra_brt}
    }
    \hfill
    \subfigure[Monthly anomaly counts in Cassandra.]{
        \includegraphics[width=0.3\textwidth]{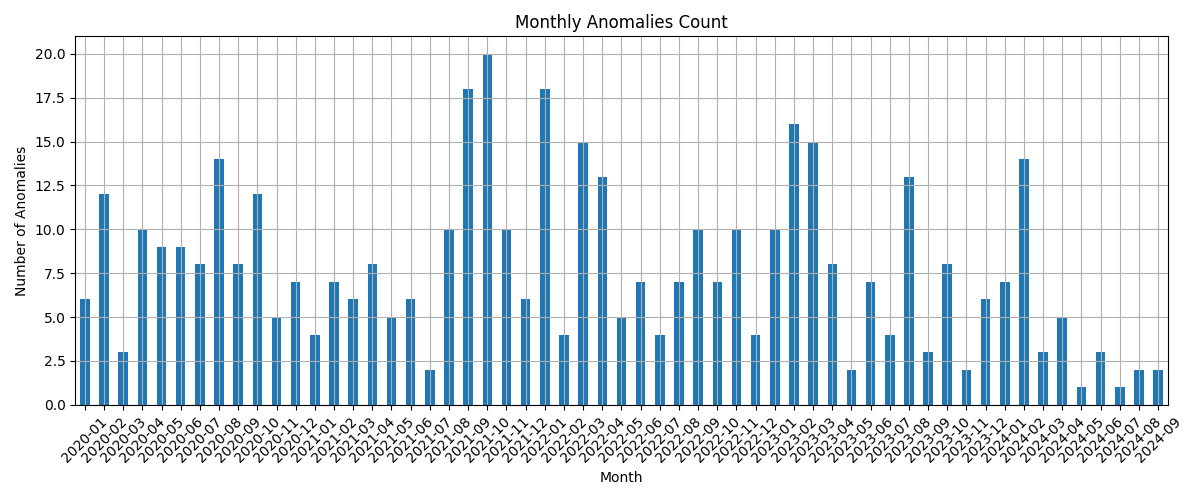}
        \label{fig:cassandra_mac}
    }
    \hfill
    \subfigure[TF-IDF clustering of anomalous summaries in Cassandra.]{
        \includegraphics[width=0.3\textwidth]{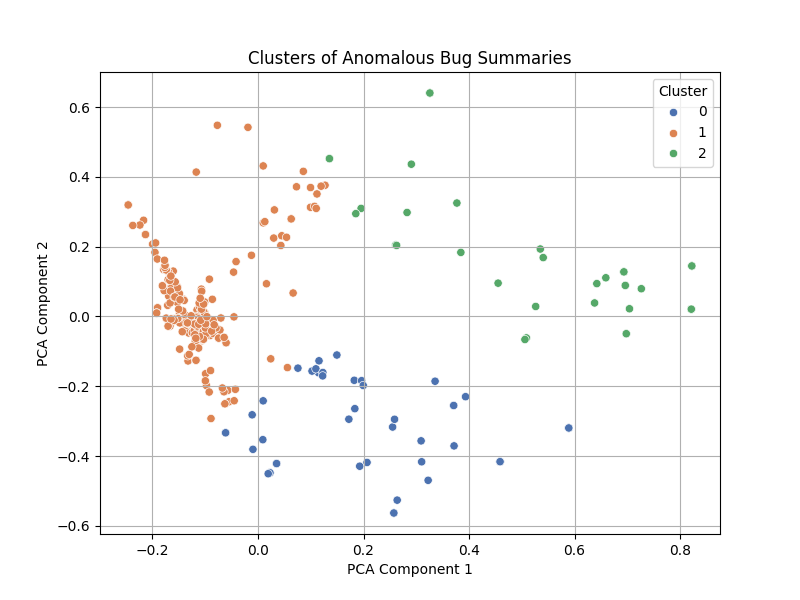}
        \label{fig:cassandra_clusters}
    }
    \caption{Anomaly detection and clustering visualizations for the Cassandra project.}
    \label{fig:cassandra_combined}
\end{figure*}

\subsection{Firefox}
In \textit{Firefox}, anomalies were persistent and wide-ranging. Clustered summaries revealed:
\begin{itemize}
    \item \textbf{Cluster 0:} UI/UX issues — \textit{tabs}, \textit{windows}, \textit{file}.
    \item \textbf{Cluster 1:} Test automation and components — \textit{browser}, \textit{intermittent}.
    \item \textbf{Cluster 2:} Navigation and layout — \textit{page}, \textit{tab}, \textit{bar}.
\end{itemize}

\begin{figure*}[ht]
    \centering
    \subfigure[Bug resolution times in Firefox (anomalies in red).]{
        \includegraphics[width=0.3\textwidth]{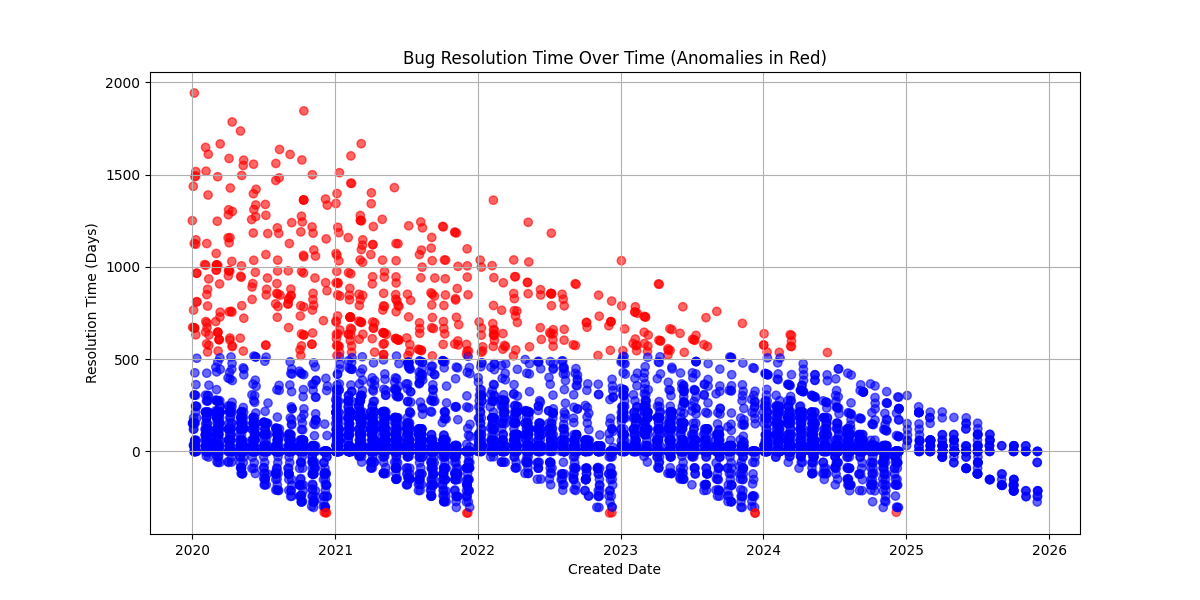}
        \label{fig:firefox_brt}
    }
    \hfill
    \subfigure[Monthly anomaly counts in Firefox.]{
        \includegraphics[width=0.3\textwidth]{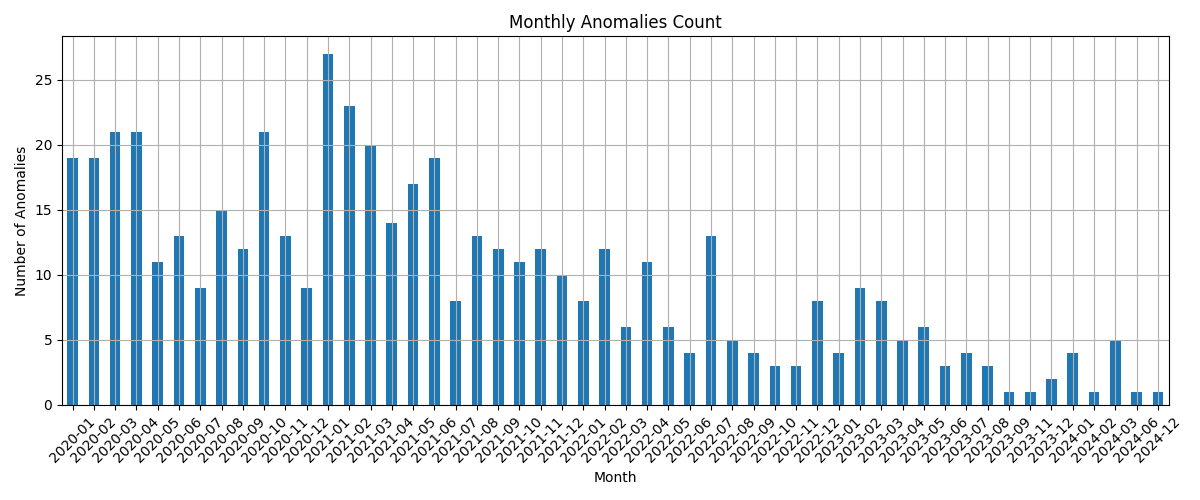}
        \label{fig:firefox_mac}
    }
    \hfill
    \subfigure[Clustering of anomalous summaries in Firefox.]{
        \includegraphics[width=0.3\textwidth]{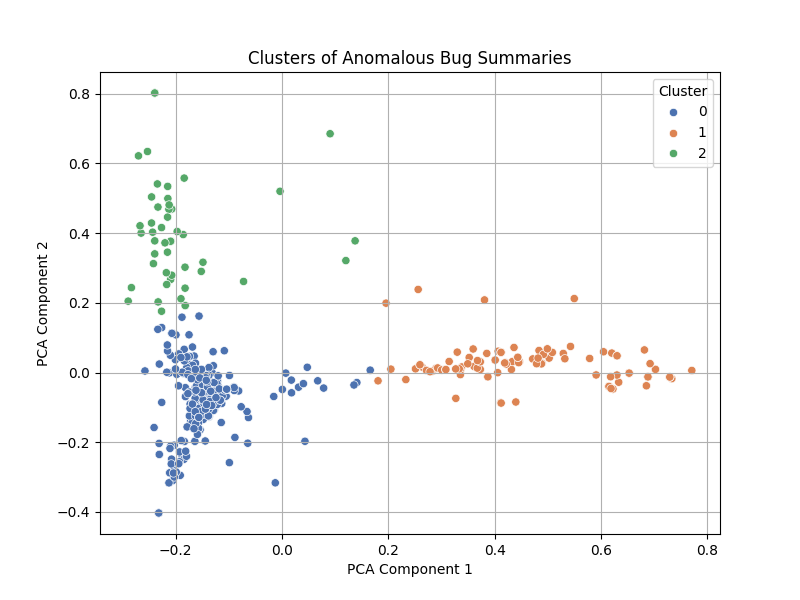}
        \label{fig:firefox_clusters}
    }
    \caption{Anomaly detection and clustering visualizations for the Firefox project.}
    \label{fig:firefox_combined}
\end{figure*}

\subsection{Hadoop}
The \textit{Hadoop} project exhibited persistent anomaly spikes early in the timeline. Keyword themes included:
\begin{itemize}
    \item \textbf{Cluster 0:} ABFS-specific bugs — \textit{abfs}, \textit{token}, \textit{header}.
    \item \textbf{Cluster 1:} S3 and generic fixes — \textit{s3a}, \textit{fix}, \textit{update}.
    \item \textbf{Cluster 2:} Upgrades and library dependencies — \textit{upgrade}, \textit{commons}.
\end{itemize}

\begin{figure*}[ht]
    \centering
    \subfigure[Bug resolution anomalies in Hadoop.]{
        \includegraphics[width=0.3\textwidth]{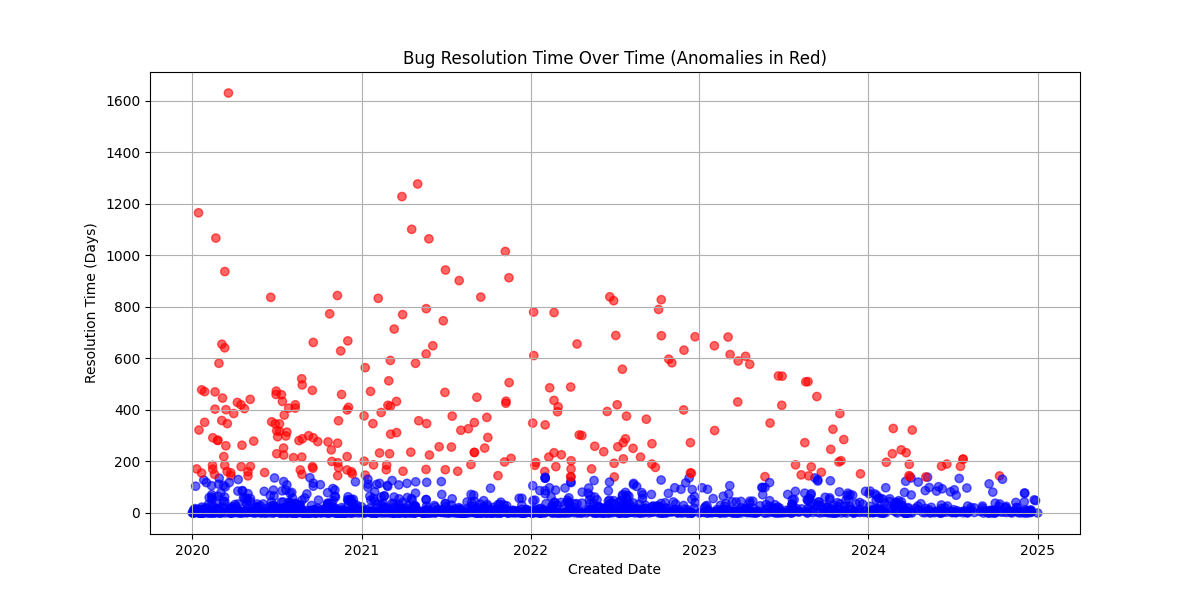}
        \label{fig:hadoop_brt}
    }
    \hfill
    \subfigure[Monthly anomaly count in Hadoop.]{
        \includegraphics[width=0.3\textwidth]{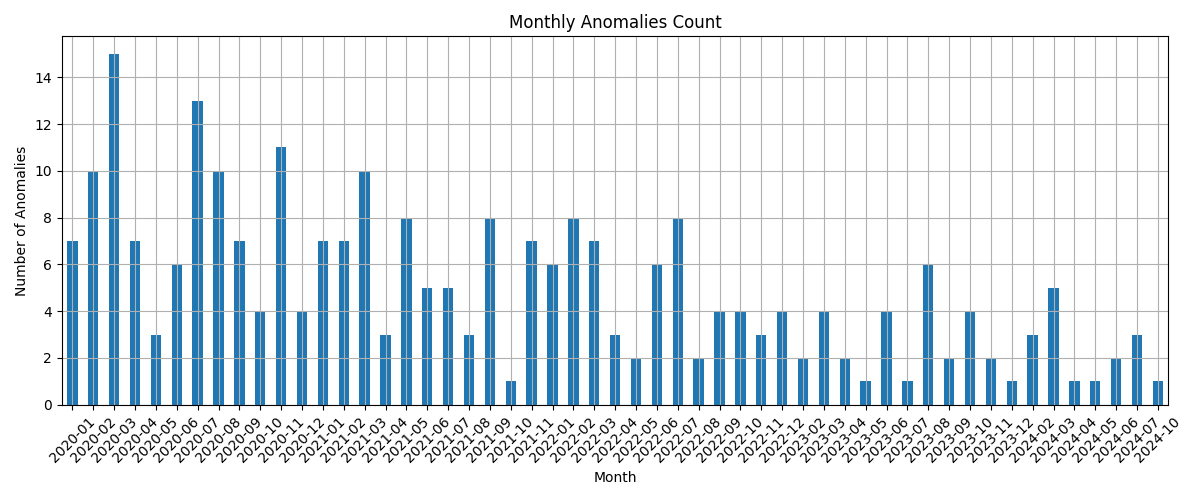}
        \label{fig:hadoop_mac}
    }
    \hfill
    \subfigure[Anomalous summary clusters in Hadoop.]{
        \includegraphics[width=0.3\textwidth]{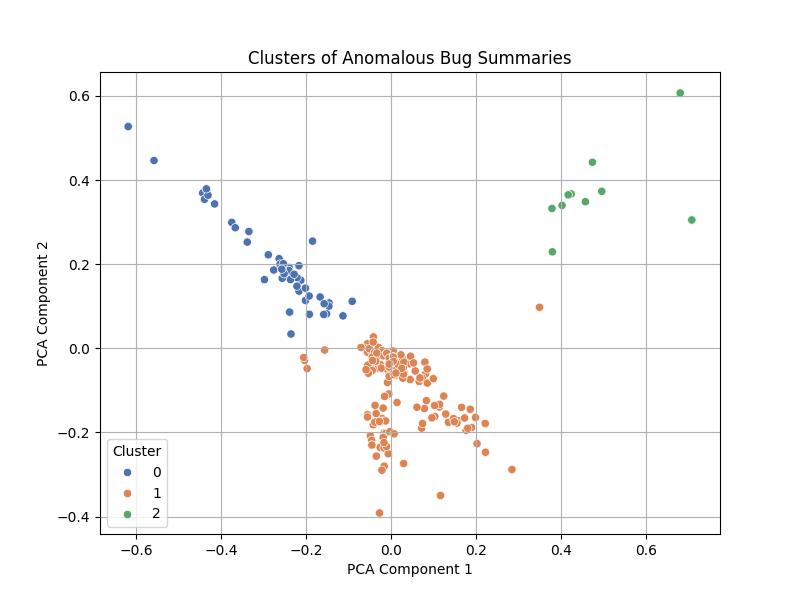}
        \label{fig:hadoop_clusters}
    }
    \caption{Anomaly detection and clustering visualizations for the Hadoop project.}
    \label{fig:hadoop_combined}
\end{figure*}

\subsection{HBase}
\textit{HBase} showed moderate anomaly presence. Clustering revealed:
\begin{itemize}
    \item \textbf{Cluster 0:} Metrics and compatibility — \textit{jdk17}, \textit{metrics}.
    \item \textbf{Cluster 1:} Table and WAL-related terms.
    \item \textbf{Cluster 2:} Backport and connector-related bugs.
\end{itemize}

\begin{figure*}[ht]
    \centering
    \subfigure[Bug resolution anomalies in HBase.]{
        \includegraphics[width=0.3\textwidth]{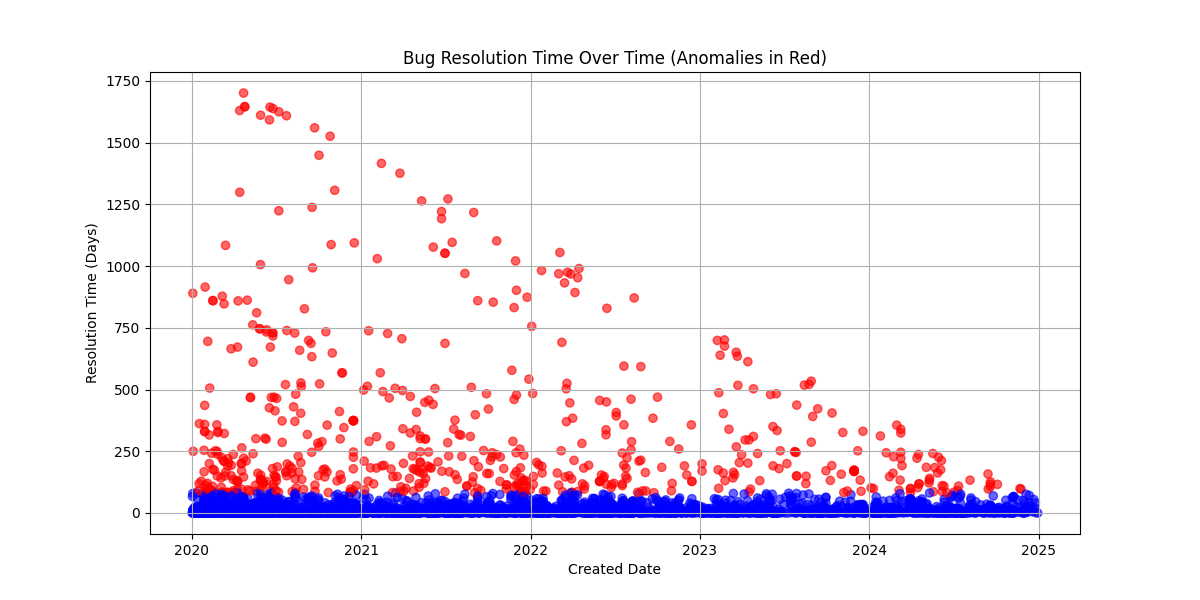}
        \label{fig:hbase_brt}
    }
    \hfill
    \subfigure[Monthly anomaly count in HBase.]{
        \includegraphics[width=0.3\textwidth]{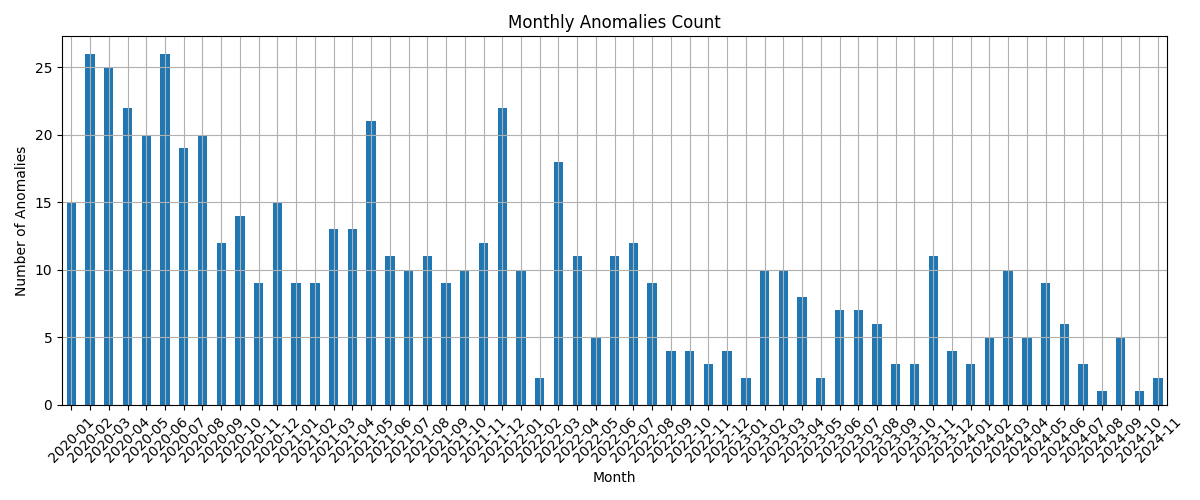}
        \label{fig:hbase_mac}
    }
    \hfill
    \subfigure[Anomalous summary clusters in HBase.]{
        \includegraphics[width=0.3\textwidth]{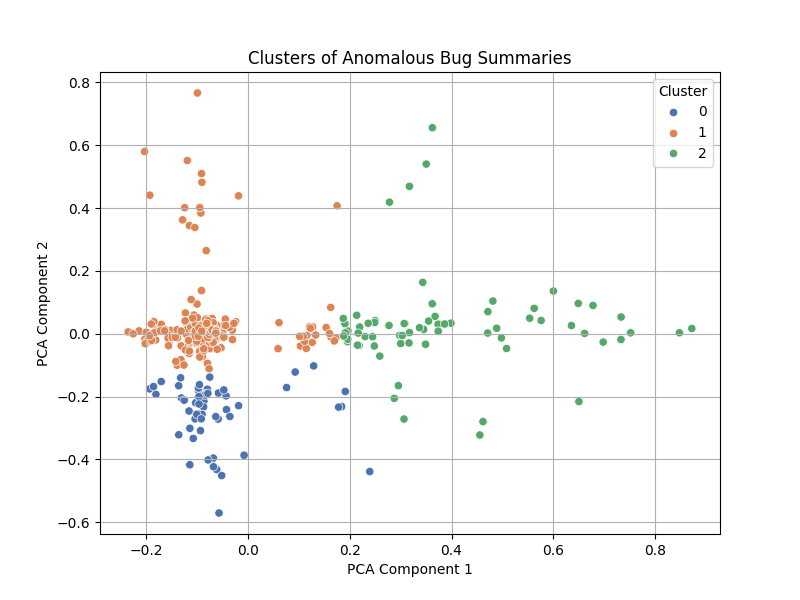}
        \label{fig:hbase_clusters}
    }
    \caption{Anomaly detection and clustering visualizations for the HBase project.}
    \label{fig:hbase_combined}
\end{figure*}

\subsection{SeaMonkey}
\textit{SeaMonkey} had relatively few long-duration anomalies. Its clusters included:
\begin{itemize}
    \item \textbf{Cluster 0:} Application behavior and usability.
    \item \textbf{Cluster 1:} Help and documentation errors.
    \item \textbf{Cluster 2:} Shutdown and table merge bugs.
\end{itemize}

% \begin{figure*}[ht]
%     \centering
%     \subfigure[Bug resolution anomalies in SeaMonkey.]{
%         \includegraphics[width=0.3\textwidth]{graphs/seamonkey_brt.png}
%         \label{fig:seamonkey_brt}
%     }
%     \hfill
%     \subfigure[Monthly anomaly count in SeaMonkey.]{
%         \includegraphics[width=0.3\textwidth]{graphs/seamonkey_mac.png}
%         \label{fig:seamonkey_mac}
%     }
%     \hfill
%     \subfigure[Anomalous summary clusters in SeaMonkey.]{
%         \includegraphics[width=0.3\textwidth]{graphs/seamonkey_clusters.png}
%         \label{fig:seamonkey_clusters}
%     }
%     \caption{Anomaly detection and clustering visualizations for the SeaMonkey project.}
%     \label{fig:seamonkey_combined}
% \end{figure*}

\subsection{Spark}
\textit{Spark} had the most consistent anomaly density over time. Summary clusters included:
\begin{itemize}
    \item \textbf{Cluster 0:} API/connectivity — \textit{connect}, \textit{python}.
    \item \textbf{Cluster 1:} PySpark errors and upgrades.
    \item \textbf{Cluster 2:} SQL/function-level enhancements.
\end{itemize}

% \begin{figure*}[ht]
%     \centering
%     \subfigure[Bug resolution anomalies in Spark.]{
%         \includegraphics[width=0.3\textwidth]{graphs/spark_brt.png}
%         \label{fig:spark_brt}
%     }
%     \hfill
%     \subfigure[Monthly anomaly count in Spark.]{
%         \includegraphics[width=0.3\textwidth]{graphs/spark_mac.png}
%         \label{fig:spark_mac}
%     }
%     \hfill
%     \subfigure[Summary clustering of anomalies in Spark.]{
%         \includegraphics[width=0.3\textwidth]{graphs/spark_clusters.png}
%         \label{fig:spark_clusters}
%     }
%     \caption{Anomaly detection and clustering visualizations for the Spark project.}
%     \label{fig:spark_combined}
% \end{figure*}

\subsection{Thunderbird}
The \textit{Thunderbird} dataset showed both interface and backend issues:
\begin{itemize}
    \item \textbf{Cluster 0:} Message/folder management — \textit{window}, \textit{folder}.
    \item \textbf{Cluster 1:} JavaScript/intermittent tests.
    \item \textbf{Cluster 2:} IMAP account bugs — \textit{doesn}, \textit{opening}.
\end{itemize}

% \begin{figure*}[ht]
%     \centering
%     \subfigure[Bug resolution anomalies in Thunderbird.]{
%         \includegraphics[width=0.3\textwidth]{graphs/thunderbird_brt.png}
%         \label{fig:thunderbird_brt}
%     }
%     \hfill
%     \subfigure[Monthly anomaly count in Thunderbird.]{
%         \includegraphics[width=0.3\textwidth]{graphs/thunderbird_mac.png}
%         \label{fig:thunderbird_mac}
%     }
%     \hfill
%     \subfigure[Summary clustering of anomalies in Thunderbird.]{
%         \includegraphics[width=0.3\textwidth]{graphs/thunderbird_clusters.png}
%         \label{fig:thunderbird_clusters}
%     }
%     \caption{Anomaly detection and clustering visualizations for the Thunderbird project.}
%     \label{fig:thunderbird_combined}
% \end{figure*}

\subsection{Cross-Repository Observations}
Across repositories, several trends emerged:
\begin{itemize}
    \item Most anomalies are skewed early in the project timeline.
    \item Projects with high CI/CD adoption (e.g., Spark, Firefox) exhibited frequent test-related anomalies.
    \item Summary clustering consistently separated anomalies into UI/UX, backend logic, and infrastructure themes.
\end{itemize}
These patterns suggest systemic causes of prolonged bug resolution, including flaky test infrastructure, backlog accumulation, and under-maintained modules.

\begin{table}[ht]
\centering
\caption{Summary of Anomaly Clusters Across Software Projects}
\label{tab:anomaly_summary}
\small
\begin{tabular}{@{} ll p{5cm} @{}}
\toprule
\textbf{Project} & \textbf{Cluster} & \textbf{Themes/Keywords} \\
\midrule
\multirow{3}{*}{Cassandra} 
 & Cluster 0 & Test instability (test, failure, flaky) \\
 & Cluster 1 & Enhancement proposals (CEP, add, update) \\
 & Cluster 2 & Core packages (apache, org) \\
\addlinespace
\midrule
\multirow{3}{*}{Firefox} 
 & Cluster 0 & UI/UX issues (tabs, windows, file) \\
 & Cluster 1 & Test automation and components \\
 & Cluster 2 & Navigation and layout (page, tab, bar) \\
\addlinespace
\midrule
\multirow{3}{*}{Hadoop} 
 & Cluster 0 & ABFS-specific bugs (abfs, token, header) \\
 & Cluster 1 & S3 and generic fixes (s3a, fix, update) \\
 & Cluster 2 & Upgrades and dependencies\\
\addlinespace
\midrule
\multirow{3}{*}{HBase} 
 & Cluster 0 & Metrics and compatibility (jdk17, metrics) \\
 & Cluster 1 & Table and WAL-related terms \\
 & Cluster 2 & Backport and connector-related bugs \\
\addlinespace
\midrule
\multirow{3}{*}{SeaMonkey} 
 & Cluster 0 & Application behavior and usability \\
 & Cluster 1 & Help and documentation errors \\
 & Cluster 2 & Shutdown and table merge bugs \\
\addlinespace
\midrule
\multirow{3}{*}{Spark} 
 & Cluster 0 & API/connectivity (connect, python) \\
 & Cluster 1 & PySpark errors and upgrades \\
 & Cluster 2 & SQL/function-level enhancements \\
\addlinespace
\midrule
\multirow{3}{*}{Thunderbird} 
 & Cluster 0 & Message/folder management (window, folder) \\
 & Cluster 1 & JavaScript/intermittent tests \\
 & Cluster 2 & IMAP account bugs (doesn, opening) \\
% \midrule

% \multicolumn{3}{@{}p{\textwidth}@{}}{\textbf{Cross-Repository Trends:}
% \begin{itemize}[leftmargin=*,nosep]
% \item Early-project skew in anomaly occurrences
% \item Frequent test-related anomalies in CI/CD-heavy projects (e.g., Spark, Firefox)
% \item Common cluster themes: UI/UX, backend logic, infrastructure
% \item Prolonged resolution linked to flaky tests, backlog, under-maintained modules
% \end{itemize}} \\
\bottomrule
\end{tabular}
\end{table}

\section{Discussion}

The results across seven open-source repositories demonstrate that anomalous bug resolution times are not only prevalent but also exhibit a distinct thematic structure. This section discusses key observations, implications for project maintainers, and limitations of our approach.

\subsection{Systemic Patterns in Resolution Anomalies}

Across repositories such as \textit{Cassandra}, \textit{Spark}, and \textit{Hadoop}, anomalies consistently appeared earlier in the project timelines. This suggests the presence of technical debt and backlog accumulation in earlier phases of project evolution. In contrast, repositories like \textit{SeaMonkey} and \textit{HBase} exhibited fewer anomalies, potentially reflecting different triage practices or lower issue volume.

\subsection{Themes Behind Long-Standing Bugs}

Thematic clustering of anomalous bug summaries reveals a repeating triad:
\begin{enumerate}
    \item \textbf{Test Failures and Instability:} Particularly in repositories like \textit{Firefox} and \textit{Cassandra}, a large share of anomalies involve flaky or intermittent tests.
    \item \textbf{Infrastructure and Dependencies:} Bugs involving API support, configuration, or external dependencies (e.g., \textit{s3a}, \textit{abfs}, \textit{connect}) consistently formed separate clusters.
    \item \textbf{User Interface Issues:} UI bugs—such as tab behavior, layout issues, or window inconsistencies—also appeared in delayed resolutions, especially in \textit{Firefox} and \textit{Thunderbird}.
\end{enumerate}

These themes suggest that the nature of a bug significantly influences how quickly it can be resolved, with infrastructure and testing issues likely requiring coordination across subsystems or environments.

\subsection{Implications for Software Maintenance}

Understanding the causes and types of anomalous bug reports can inform:
\begin{itemize}
    \item \textbf{Triage Prioritization:} Projects may consider flagging known anomaly patterns (e.g., flaky test failures) early in the triage process.
    \item \textbf{Process Improvement:} Persistent long-tail resolution patterns may indicate the need for improved backlog management, test infrastructure, or contributor engagement.
    \item \textbf{Tooling Opportunities:} Anomaly detection can be integrated into issue trackers to flag bugs at risk of becoming long-standing, based on similarity to known patterns.
\end{itemize}

\subsection{Limitations and Future Work}

The granularity of available metadata limits our approach. For instance, we did not consider developer-assigned labels or linked commits. Additionally, clustering is sensitive to the choice of $k$, and while $k=3$ provided interpretable results, alternative models (e.g., hierarchical clustering, topic models) may uncover deeper structures.

Future work may explore:
\begin{itemize}
    \item Integration of additional features such as assignee history, code churn, or number of comments.
    \item Longitudinal tracking of anomaly rates as a quality metric.
    \item Predictive modeling of bugs at risk of becoming anomalous based on early metadata signals.
\end{itemize}

\section{Conclusion}

This paper presented a comprehensive analysis of bug resolution anomalies across seven large-scale open-source software repositories. By combining statistical outlier detection (Z-score and IQR methods) with text mining and clustering techniques (TF-IDF and KMeans), we identified not only when bugs took unusually long to resolve but also what themes these long-standing bugs shared.

Our findings reveal consistent patterns in anomaly behavior:
\begin{itemize}
    \item Anomalies tend to cluster in the early stages of project timelines, highlighting the impact of legacy technical debt.
    \item Thematic clusters across repositories consistently include flaky tests, infrastructure/API bugs, and UI-related issues.
    \item Projects with high issue throughput (e.g., Spark, Firefox) exhibit persistent resolution anomalies, suggesting the need for process scaling and prioritization tools.
\end{itemize}

These insights provide a foundation for integrating anomaly-aware analytics into issue-tracking systems. By proactively identifying bugs at risk of prolonged resolution, development teams can improve triage efficiency, focus testing resources, and reduce maintenance backlog.

Future work will explore the integration of richer features such as developer behavior, linked code changes, and social dynamics (e.g., comment activity) to enhance anomaly prediction. Additionally, applying language models to achieve a deeper semantic understanding of bug summaries could improve the quality and interpretability of clustering.

\bibliographystyle{IEEEtran}
\nocite{*}
\bibliography{ref}

\clearpage
\onecolumn
\section*{Appendix: Additional Project Visualizations}

This appendix includes anomaly detection and clustering visualizations for the remaining four projects: \textit{SeaMonkey}, \textit{Spark}, and \textit{Thunderbird}.

\begin{figure*}[ht]
    \centering
    \subfigure[Bug resolution anomalies in SeaMonkey.]{
        \includegraphics[width=0.3\textwidth]{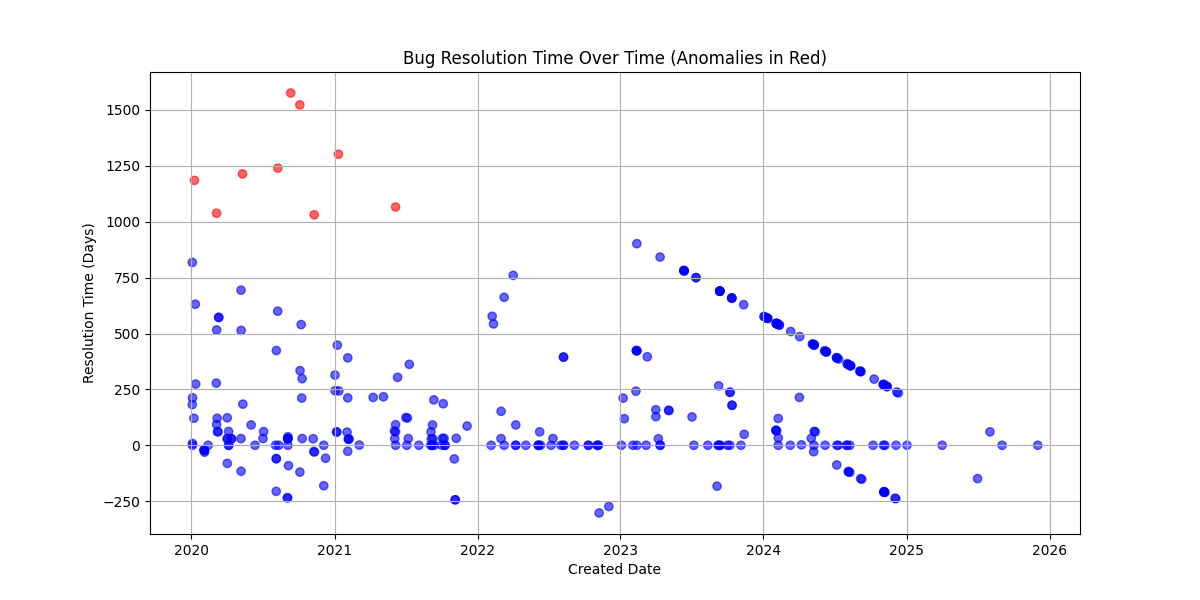}
        \label{fig:seamonkey_brt}
    }
    \hfill
    \subfigure[Monthly anomaly count in SeaMonkey.]{
        \includegraphics[width=0.3\textwidth]{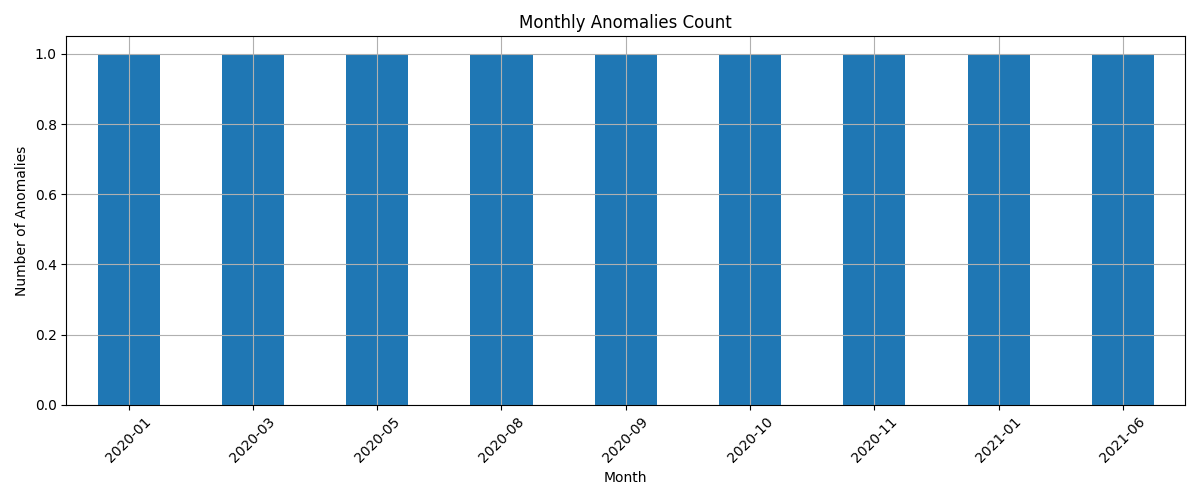}
        \label{fig:seamonkey_mac}
    }
    \hfill
    \subfigure[Anomalous summary clusters in SeaMonkey.]{
        \includegraphics[width=0.3\textwidth]{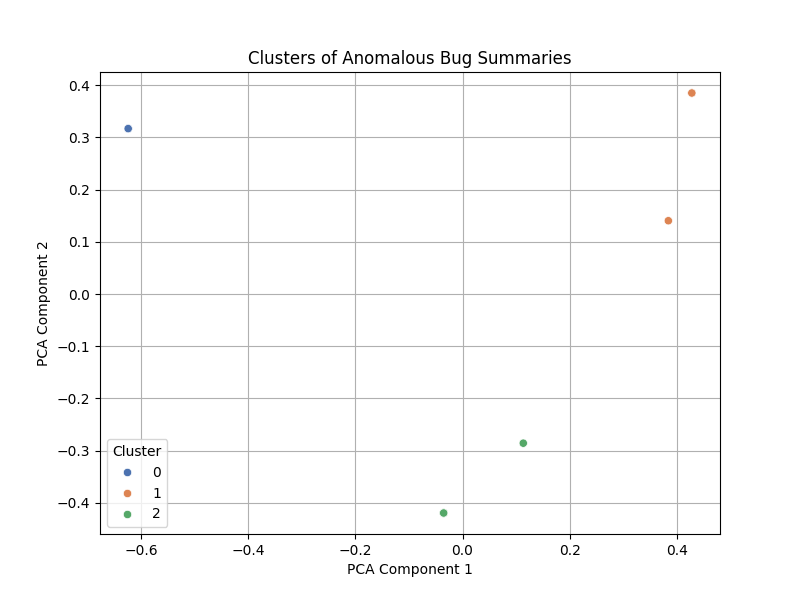}
        \label{fig:seamonkey_clusters}
    }
    \caption{Anomaly detection and clustering visualizations for the SeaMonkey project.}
    \label{fig:seamonkey_combined}
\end{figure*}

\begin{figure*}[ht]
    \centering
    \subfigure[Bug resolution anomalies in Spark.]{
        \includegraphics[width=0.3\textwidth]{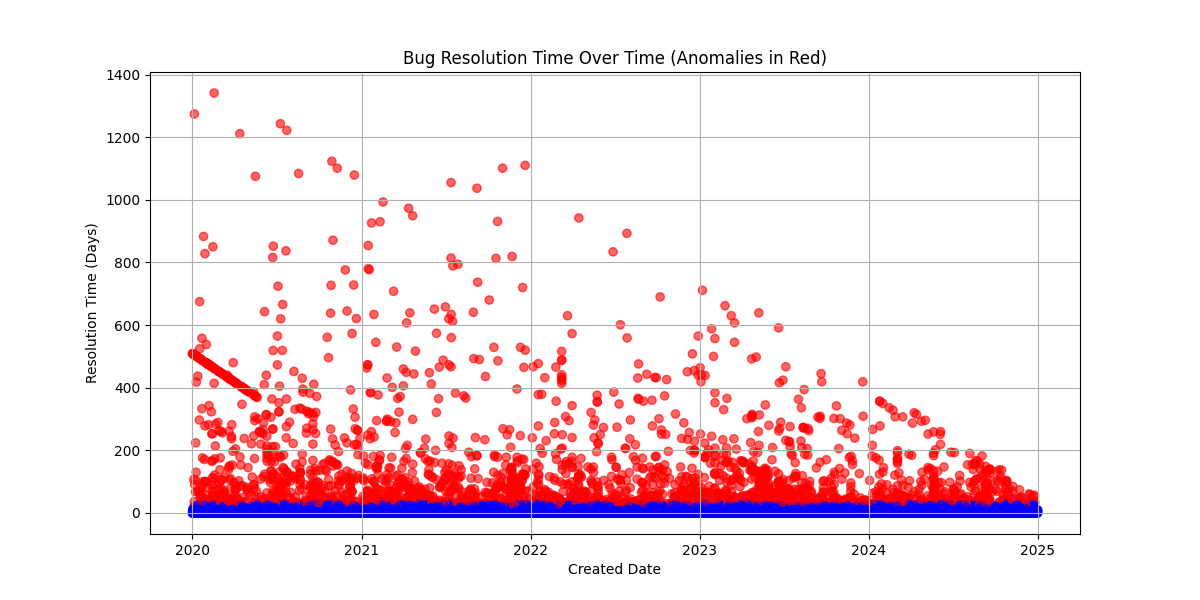}
        \label{fig:spark_brt}
    }
    \hfill
    \subfigure[Monthly anomaly count in Spark.]{
        \includegraphics[width=0.3\textwidth]{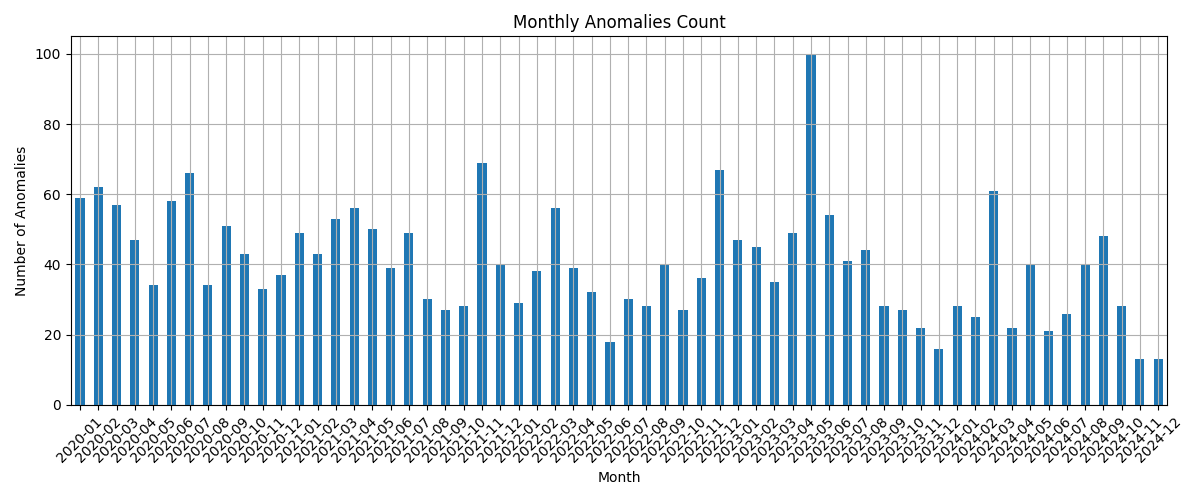}
        \label{fig:spark_mac}
    }
    \hfill
    \subfigure[Summary clustering of anomalies in Spark.]{
        \includegraphics[width=0.3\textwidth]{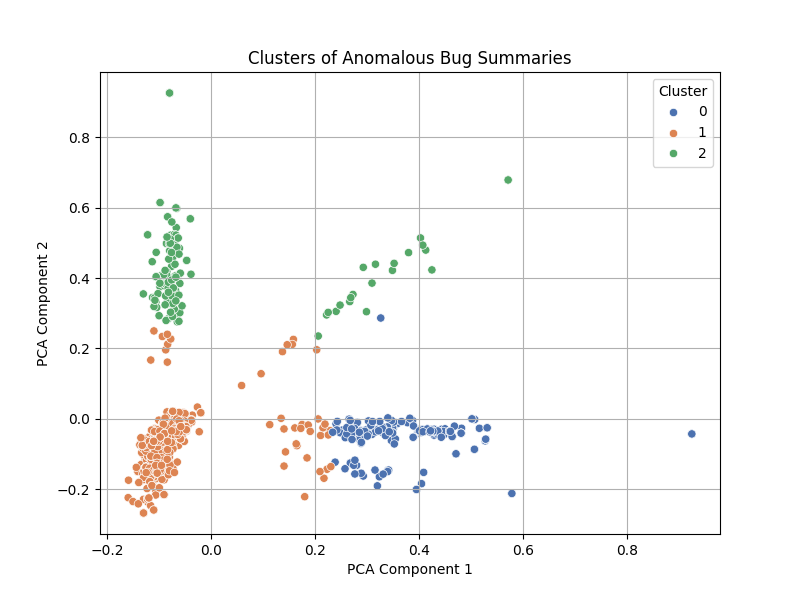}
        \label{fig:spark_clusters}
    }
    \caption{Anomaly detection and clustering visualizations for the Spark project.}
    \label{fig:spark_combined}
\end{figure*}

\begin{figure*}[ht]
    \centering
    \subfigure[Bug resolution anomalies in Thunderbird.]{
        \includegraphics[width=0.3\textwidth]{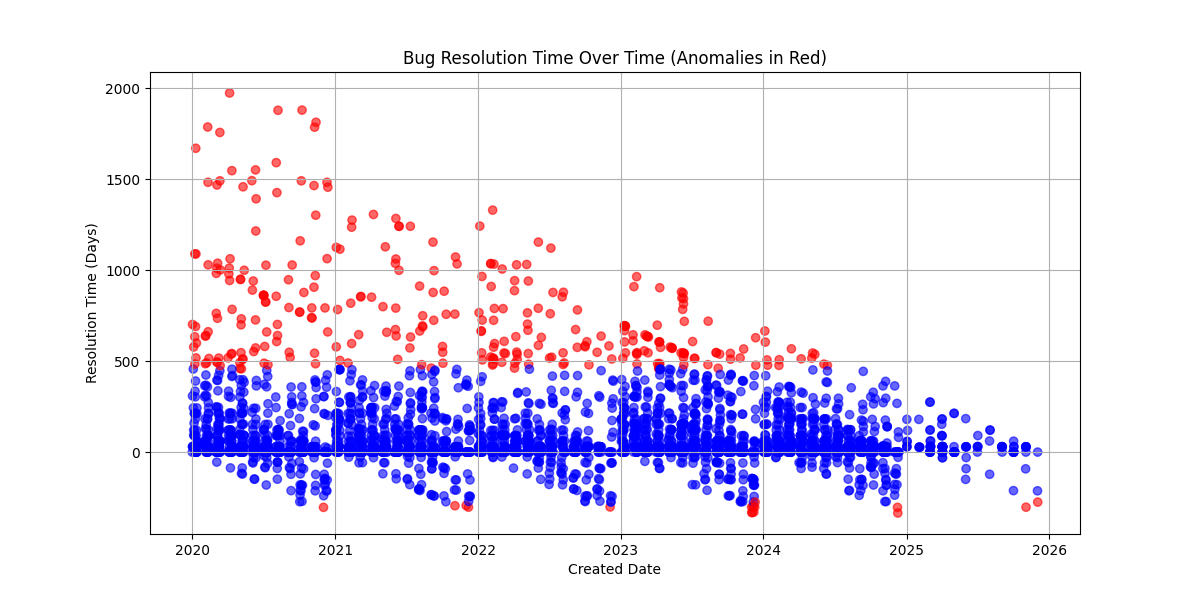}
        \label{fig:thunderbird_brt}
    }
    \hfill
    \subfigure[Monthly anomaly count in Thunderbird.]{
        \includegraphics[width=0.3\textwidth]{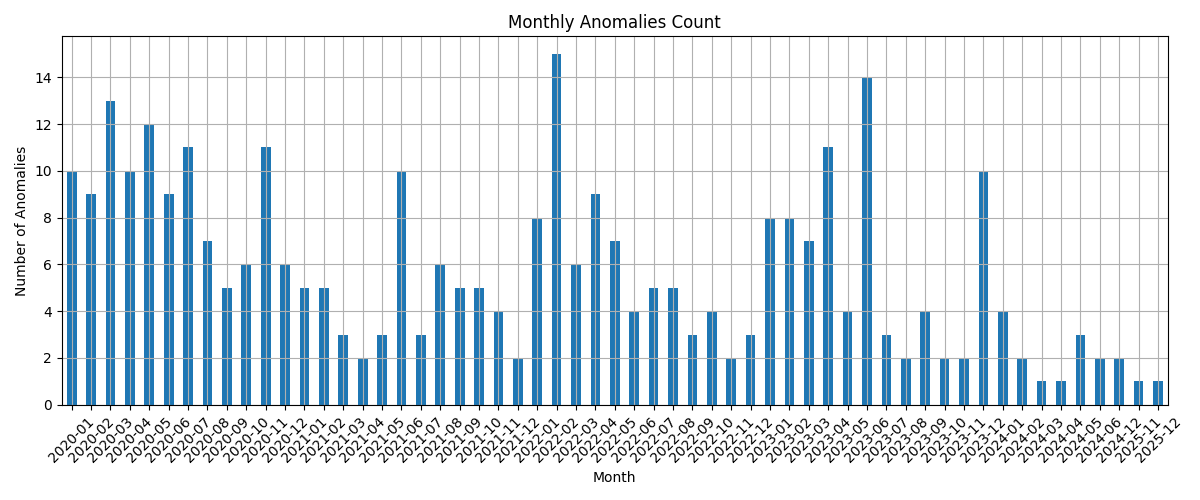}
        \label{fig:thunderbird_mac}
    }
    \hfill
    \subfigure[Summary clustering of anomalies in Thunderbird.]{
        \includegraphics[width=0.3\textwidth]{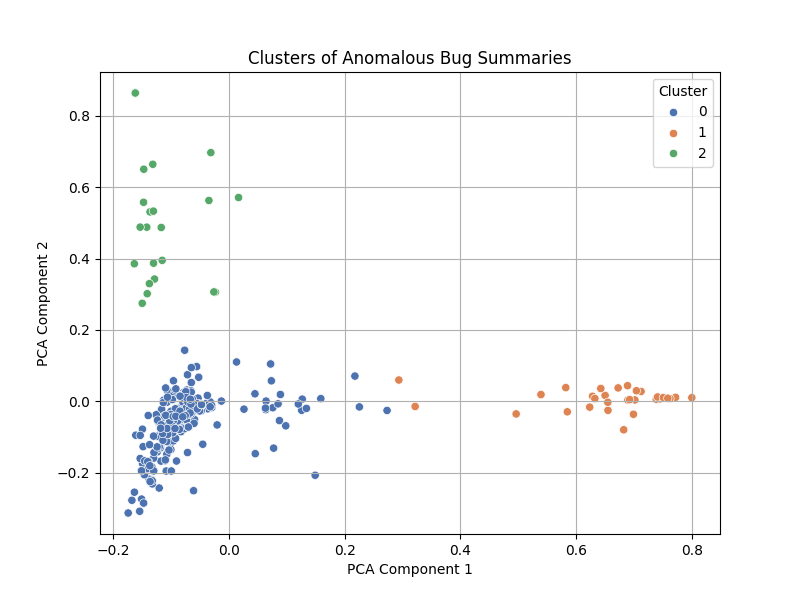}
        \label{fig:thunderbird_clusters}
    }
    \caption{Anomaly detection and clustering visualizations for the Thunderbird project.}
    \label{fig:thunderbird_combined}
\end{figure*}

\end{document}